\documentclass[pra,floatfix,preprint,nofootinbib,eqsecnum,12pt]{revtex4}
\usepackage{epsfig}
\usepackage{bm}
\usepackage{amsmath}
\input epsf
\numberwithin{equation}{section}

\def\Hc{{\cal H}}
\def\HP{$(H,|\Psi\rangle)$}
\def\Rc{{\cal R}}
\def\Re{\text{Re}}
\def\ab{{\bar\alpha}}
\def\bb{{\bar\beta}}

\def\be{\begin{equation}}
\def\ee{\end{equation}}

\begin{document}
\vspace{1cm}

\title{Decoherent Histories Quantum Mechanics \\  Starting with Records of What Happens}
\author{James B.~Hartle}

\email{hartle@physics.ucsb.edu}

\affiliation{Santa Fe Institute, Santa Fe, NM 87501}
\affiliation{Department of Physics, University of California,Santa Barbara, CA 93106-9530}

\date{\today }

\begin{abstract}
We present a formulation of the decoherent (or consistent) histories quantum theory of closed systems starting with records of what histories happen. Alternative routes to a formulation  of quantum theory like this one can  be useful both for understanding quantum mechanics  and for generalizing and extending it to new realms of application and experimental test. 
\end{abstract}



\vspace{ .2in}

\maketitle

\bibliographystyle{unsrt}


\section{Introduction}
\label{intro}

Decoherent (or consistent) histories\footnote{For classic expositions, some by founders of the subject, see \cite{classicDH}. For a tutorial see \cite{Har93a}. } quantum mechanics  (DH) is logically consistent, consistent with experiment as far as is known, consistent with the rest of modern physics such as special relativity and quantum field theory, general enough or cosmology, and generalizable to apply to semiclassical quantum spacetime.  It may not be the only quantum framework with these properties but it is the only one we have at present. 

Other formulations of quantum mechanics applicable in more restricted circumstances can be seen as restrictions of DH \cite{Har05}.   For example, the Copenhagen quantum mechanics of measurement outcomes (CQM) can be seen as an approximation to DH that is appropriate when some part of  the closed system is a measurement situation (e.g. \cite{Har91a}). 

DH can be formulated in ways that are different from the traditional one expounded in \cite{classicDH}. An example is the formulation in terms of extended probabilities \cite{exprob}. This paper presents yet  another alternative formulation   starting from records of histories that  requires no extension of usual probability theory.

Alternative formulations of DH are of interest for at least two reasons: 

(1) To have different starting points for looking for theories that are close to DH on the scales on which it has been tested and applied, but  which make different experimentally testable  predictions on scales where it has not yet been tested. In quantum cosmology DH is applied the whole universe --- an enormous extrapolation of scale from those on which its principles could be claim to be tested.  Quantum theories that agree with DH on the scales which it has been tested,  but make different predictions on cosmological scales would be of great interest. 

(2) The second reason that alternative formulations of DH are useful is to facilitate exposition. It is sometimes said that understanding an idea, concept, or thing means being able to describe it in different ways. Different formulations emphasize different aspects of the theory and different scientists may find the theory more accessible, useful,  or more in conformity with their prejudices in one formulation than another.  The Lagrangian and Hamiltonian formulations of classical mechanics and the corresponding sum-over-histories (Lagrangian)  and Hamiltonian formulations of usual quantum theory are examples of how different formulations of the same theory can be useful. 

The most general objective of any  quantum mechanics of a closed system like the universe  are the probabilities for the individual members of sets of alternative coarse-grained histories of the system. According to quantum theory the Moon might take any orbit in its progress around the Earth.  But, in its present situation, the quantum mechanical probability for suitably coarse grained alternative histories of the Moon's  center of mass position are vastly larger for an orbit conforming to Newton's laws of motion than any other. Here, `coarse-grained'  means following the center of mass position not at every time but only at a sequence of times, and these not  to arbitrary precision but with suitable imprecision. These probabilities are instructions for betting on the outcome of observations of the Moon's  orbit \cite{deF37}.  In ordinary parlance we would say these probabilities are for which orbit happens, or occurs, or is realized. 

To construct a theory of such probabilities it would seem natural to first give a quantum mechanical meaning to what happens and then define probabilities that allow us to bet on what happens. A theory is successful if bettors using those probabilities win.  That is the approach we will follow in this formulation of quantum mechanics. We  sketch the argument as follows:

We will say a history of events `happens' if there is a record of the history at one time according to a specific non-probabilistic measure of correlation between record and history\footnote{We will have more to say about this usage of `happen' in Section \ref{happened}.}.  We assume that probabilities for records are given by Born's rule\footnote{Thus we are not attempting here to derive Born's  rule from some other assumption. }. The correlation between record and history means that this is the probability for the history to `happen'. The result is a formulation of quantum theory that is more general than DH. We show how DH is recovered with a suitably strong notion of record. 

The paper is structured as follows.  Section \ref{model} introduces a model quantum universe in a box that along with necessary notation to describe sets of alternative histories.  Section \ref{RHist} introduces the correlation function that defines the notions of a record of a history, a recorded set of alternative histories, and defines probabilities for the individual histories in a recorded set. Section \ref{happened} gives a  discussion of the use of the word `happen' in quantum mechanics. Section \ref{comparison} compares three different formulations of DH.
Section \ref{generalizations} describes the motivations for seeking generalizations of DH and the utility of different formulations of it for doing that. 
There is a very brief conclusion in Section \ref{conclusion}


\section{A Model Quantum Universe}
\label{model}

\begin{figure}[t]
\includegraphics[width=3in]{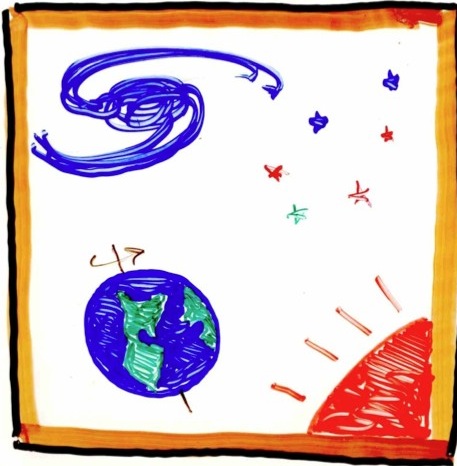}\hfill
\caption{A simple model of a closed quantum system is a universe of quantum matter fields inside a large closed box (say, 20,000 Mpc on a side) with fixed flat spacetime inside. Everything is a physical system inside the box --- galaxies, stars, planets, human beings, observers and observed, measured and measuring. The most general objectives for prediction are the probabilities of the individual members of decoherent sets of alternative coarse grained histories that describe what goes on in the box. That includes histories describing any measurements that take place there. There is no observation or other intervention from outside.  }
\label{box}
\end{figure}

To keep the discussion manageable, we consider a closed quantum system in the approximation that gross quantum fluctuations in the geometry of spacetime
can be neglected. The closed system can then be thought of as a large (say
$\gtrsim$ 20,000 Mpc), perhaps expanding, box of particles and fields in a 
fixed, flat, background spacetime (Figure \ref{box}). There is  a well defined notion of time in any particular Lorentz frame. The familiar apparatus of textbook quantum mechanics then applies ---  a Hilbert space $\Hc$, operators, states,  and their unitary evolution.

The important thing is that everything is contained within the box, in particular galaxies, planets, observers and
observed, measured subsystems, and any apparatus that measures
them.  This is a model cosmology and the most general physical context for prediction.

The basic theoretical inputs for prediction are the  Hamiltonian $H$ governing evolution and quantum state of the box $|\Psi\rangle$, written in the Heisenberg picture for convenience, and assumed pure for simplicity. Input theory is  then \HP.
We use the theory to predict probabilities for which of a set of alternative coarse-grained histories of the contents of the box happens. 

The simplest notion of a set of histories is described by giving a sequence of  yes-no alternatives at a series of times. 
For example: Is a particle in this region of the box at this time --- yes or no? Such alternatives at one moment of time are described by an exhaustive set of exclusive Heisenberg picture projection operators $\{P_\alpha(t)\}$, $\alpha=1,2,3.\cdots$ acting in $\Hc$. These satisfy 
\be
\label{projections1}
P_\alpha(t)P_{\alpha'}(t) = \delta_{\alpha\alpha'} P_\alpha(t), \quad \sum_\alpha P_\alpha(t) = I . 
\ee

Projection operators representing the same quantity at different times are connected by unitary evolution by the Hamiltonian $H$
\be
P_\alpha(t') = e^{iH(t'-t)}P_\alpha(t) e^{-iH(t'-t)} . 
\label{un-evol}
\ee

A set of alternative coarse-grained histories is specified by a sequence of such sets at a series of times $t_1,t_2, \cdots t_n$. An individual history corresponds to a particular sequence of events  $\alpha \equiv (\alpha_1,\alpha_2, \cdots, \alpha_n)$ and is represented by the corresponding chain of projections:
\be
C_\alpha \equiv P^n_{\alpha_n}(t_n) \dots P^1_{\alpha_1}(t_1) .
\label{class-closed}
\ee
An immediate consequence of this and \eqref{projections1} is that 
\be
\sum_\alpha C_\alpha =I \ ,
\label{exhaustC}
\ee
showing that the set of histories is exhaustive. 


This description of histories is  analogous to those for sequences of ideal measurements in Copnhagen quantum theory (e.g. \cite{Gro52}). However, there are at least two crucial differences. First, there is no posited separate classical world as in CQM. It's all quantum. Second, the alternatives represented by the $P$'s are not restricted to measurement outcomes. They might, for example, refer to the orbit to the Moon when no one is looking at it, or to the magnitude of density  fluctuations in the early universe when there were neither observers nor apparatus to measure them. Laboratory measurements can of course be described in terms of correlations between two particular kinds of  subsystems of the universe --- one being measured the other doing the measuring. But laboratory measurements play no central role in formulating DH, and are just a small part of what it can predict\footnote{Indeed, the Copenhagen quantum mechanics (CQM) of measured subsystems is an approximation appropriate for measurement situations to the more general quantum mechanics of closed systems. See, e.g. \cite{Har91a} Section II.10.}. This is a model cosmology.


\section{Recorded Histories}
\label{RHist}

\subsection{Records}
\label{records} 

A set of alternatives histories $\{C_\alpha\}$ extending over time is  {\it recorded} if there is a set of alternatives at one time  --- the records --- that are highly correlated with the histories in the state $|\Psi\rangle$.  The records are represented by a set of projection operators  $\{ R_\alpha\}$ satisfying \eqref{projections1}.  It is natural to take the time of the records $t_r$ to be after the final time in the histories $t_n$ to model many physical mechanisms for the creation of records\footnote{This assumption simplifies the discussion but is not necessary for the conclusions, see. e.g. \cite{GH94}.}. We  posit a correlation measure $\Rc(\alpha,\beta)$ between histories and records defined as follows:
\be
\label{meascorr}
\Rc(\alpha,\beta) \equiv \frac{1}{2}\langle\Psi | R_\alpha C_\beta + C_\beta^\dagger R_\alpha |\Psi\rangle  
=\Re \langle \Psi|R_\alpha C_\beta |\Psi\rangle. 
\ee
The measure is normalized 
\be
\label{normR}
\sum_{\alpha\beta} \Rc(\alpha,\beta) = 1
\ee
but not necessarily positive. 

We will say that $R_\alpha$ is a record of the history $C_\alpha$ when 
\be
\label{correlation}
\Rc(\alpha, \beta) \approx 0,    \quad   \alpha \ne \beta  .
\ee
When this condition is satisfied we will say that one of the histories in the set happens because there is a record of it according to the measure $\Rc$.

Realistic records of history are never exactly correlated with the history they record.  The notation $\approx$ in \eqref{correlation} indicates this. We mean by this that the off-diagonal elements vanish to an accuracy well beyond that which the records are used, or the model that computes them is defined. 

Note that the records $\{R_\alpha\}$ happen because they record themselves. Also with this definition, histories at one moment of time will be exactly recorded by themselves. 

The measure $\Rc(\alpha,\beta)$ is not a probability measure. In particular it is not restricted to positive values. Records can be both correlated and anticorrelated with the histories. This formulation of DH does not start by positing a notion of probability but rather a notion of correlation with records\footnote{In his seminal paper Everett started with a measure on alternatives which was then used to define probabilities \cite{Eve57}.}.

\subsection{Probabilities for Records and Probabilities for Histories}
\label{probability}

Probabilities can be generally understood as instructions for betting\footnote{For a concise and accessible introduction see \cite{Cavup}.}  \cite{deF37} .  To believe that the probability of an event is $p$  means that if a bookie offers you a payoff $S$ on whether the event happens you will put up a stake $pS$ and consider it a fair bet\footnote{Thus we also connect with the usual Baysian ideas if of probability, see e.g. \cite{Sre05}. }.  All the usual  rules of probability theory follow from the requirement that the bookie not be able to offer you a bet that you would always loose. You and your bookie also have to agree on how to settle a bet. Bet's are usually settled on the basis of records of what happens.  Probability is thus connected with records.  

Since they  are all at one time we can take the probabilities of the  records to be given by Born's rule\footnote{Thus we are not attempting to derive Born's rule from some other assumption.}
\be
\label{bornrule}
p_{\rm rec}(\alpha) \equiv \langle\Psi|R_\alpha|\Psi\rangle = ||R_\alpha|\Psi\rangle||^2. 
\ee
These satisfy all the usual rules of probability theory:  They are positive, sum to unity, and obey the sum rules. For example,  for two alternative histories $\alpha$ and $\beta$
\be
\label{sumrules}
p(\alpha \cup \beta) = p(\alpha) + p(\beta)
\ee
as a consequence of $R_{\alpha \cup \beta} = R_\alpha + R_\beta$. 

We then  posit that the probabilities  of the individual members of recorded set  of alternative histories $\{C_\alpha\}$ is are same as the probabilities of their records:
\be 
\label{probRH}
p_{\rm hist}(\alpha)\equiv p_{\rm rec}(\alpha)  \equiv \langle\Psi|R_\alpha|\Psi\rangle \approx \Re\langle\Psi|C_\alpha|\Psi\rangle. 
\ee
The last equality follows from summing \eqref{correlation} separately over $\alpha$ and $\beta$. 

\subsection{Fine Graining and Coarse Graining}
\label{graining}

Recorded sets of alternative histories of a closed system may divided up into families related by operations of fine and coarse graining. Consider a set of alternative histories $\{C_\alpha\}$. A coarse-graining of this set is a partition of it into larger sets of exclusive alternative histories, $\{C_\ab\}$. 
\be
\label{cg} 
\ab =\cup_{\alpha \in \ab} \alpha,   \quad      C_\ab = \sum_{\alpha \in \ab} C_\alpha .
\ee
A  coarse graining of a recorded set is again a recorded set with the records
\be
\label{cg} 
R_\ab = \sum_{\alpha \in \ab} R_\alpha
\ee
since
\be
\label{cgrecords}
\Rc(\ab,\bb) \approx  \sum_{\alpha\in\ab}\sum_{\beta\in\bb} \Re \langle\Psi|R_\alpha C_\beta|\Psi\rangle \approx \delta_{\ab\bb}.
\ee
A fine-graining of a recorded set is not necessarily recorded.

A family of recorded sets consists of recorded sets that are connected by operations of fine and coarse graining. 
Two recorded sets of histories for which there is no common recorded fine graining of which whey are both coarse grainings are said to be {\it incompatible}. 

\subsection{Independent Systems}
\label{independent}
Suppose our model universe in a box consisted of two independent, non-interacting, subsystems $a$ and $b$, with $\Hc = \Hc^a 
\otimes \Hc^b$ , $|\Psi\rangle = |\Psi^a\rangle \otimes |\Psi^b\rangle$, $H=H^a+H^b$, etc. We would expect records of histories and their probabilities to have similar decompositions\footnote{This is not automatic as shown by some formulations of quantum mechanics where it is not automatically true. Specifically we mention weak decoherence \cite{Dio04}, linear positivity \cite{GP95,Har04} and extended probabilities \cite{exprob}.}. We can check this is true as follows:

Two sets of histories $\{C_\alpha^a\}$ and $\{C_\alpha^b\}$ define a history of the whole system$\{C_\alpha^a\}$ by the relation
\begin{subequations}
\be
\label{tensorC}
C_\alpha = C_\alpha^a \otimes C_\alpha^b .
\ee
with similar relations for the records and the state
\begin{align}
\label{tensorR}
R_\alpha &= R_\alpha^a \otimes R_\alpha^b, \\
|\Psi\rangle &= |\Psi^a\rangle \otimes |\Psi^b\rangle.
\end{align}
\end{subequations}
The condition for a record of the whole system is then, from \eqref{meascorr},
\be
\label{meascorr1}
\Rc(\alpha^a,\alpha^b,\beta^a,\beta^b) 
=\Re [\langle \Psi^a |R^a_{\alpha^a} C_{\beta^a}^a |\Psi^a\rangle \langle \Psi^b |R^b_{\alpha^b} C_{\beta^b}^b |\Psi^b\rangle]\propto 
\delta_{\alpha^a\beta^a} \delta_{\alpha^b\beta^b} .
\ee
Of course, the real part of a product is not the product of the real parts. But if we sum \eqref{meascorr1} over $\alpha^b$ and  $\beta^b$ we see that $\{R^a_{\alpha^a}\}$ record the the $\{C^a_{\alpha^a}\}$ histories of subsystem $a$. Similarly the histories of subsystem $b$ are recorded.  The histories of the combined system are recorded if and only if the histories of its two independent components are.

\subsection{Strong Records and Medium Decoherence}
\label{decoherence}

In many realistic situations, where only certain variables are being followed, records of histories of those variables may be formed by interaction with variables that are being ignored  thus constituting an environment. The classic example of Joos and Zeh \cite{JZ85} is a familiar illustration:  A dust grain of millimeter size is in a superposition of positions a millimeter apart deep in intergalactic space. Roughly $10^{11}$ microwave background photons scatter from it every second. 
The characteristic time for forming records of the position of the dust grain in the scattered photons is about a nanosecond\footnote{For just a sampling of papers on records and environments beyond \cite{JZ85} see \cite{B-KZ05,RZZ13,GH13}.}.

Such situations lead to notion  of a strong recorded set of histories $\{C_\alpha\}$ for which there is a set of records $\{R_\alpha\}$ such that
\be
\label{strong} 
R_\alpha |\Psi\rangle = C_\alpha|\Psi\rangle \equiv |\Psi_\alpha\rangle . 
\ee
The $\{|\Psi_\alpha\rangle\}$ are called branch state vectors. 

Clearly \eqref{strong} implies that histories with strong records are exactly recorded. 
\be
\Rc_{st}(\alpha,\beta) = \delta_{\alpha\beta} p(\alpha) . 
\ee
which in turn implies the condition for recording \eqref{correlation} but not the other way around.

Strong records imply that the branch state vectors \eqref{strong} corresponding to different histories in the set  are mutually orthogonal
\be
\label{decoh}
\langle \Psi_\alpha | \Psi_\beta\rangle =\langle\Psi|C_\alpha^\dagger C_\beta |\Psi\rangle \approx \delta_{\alpha\beta} \ p(\alpha) .
\ee
This captures the idea of the absence of quantum interference between different histories --- decoherence.  Eq \eqref{decoh} is called the (medium) decoherence condition. A set of alternative histories satisfying \eqref{decoh} is said to decohere. 
The decoherence condition \eqref{decoh} is the starting point for the standard development of decoherent histories quantum mechanics.

In turn,  the decoherence condition \eqref{decoh} implies the existence of strong records $\{R_\alpha\}$ satisfying   \eqref{strong}. For example  we could take the projections to be $R_\alpha =|\Psi_\alpha\rangle\langle \Psi_\alpha| $.  Of course none of the records so constructed is necessarily accessible or useful to human observers. That is a much stronger and less well defined requirement. 

At this point we have completed our reformulation of DH by starting with recorded histories.  DH is RH with strong records.  The rest is commentary. 


\section{What Happened?}
\label{happened}

The Introduction motivated the recorded histories  formulation of DH by saying that we would {\it first} define what happened and {\it then} posit probabilities for betting on what happened. This section elaborates this usage of `happen' and its limitations. 

\subsection{Inferring the  Past from Present Data}
\label{pd}

What happened in the past is central to many areas of science --- cosmology, geology, planetary science,  evolutionary biology, and human history to name a few.  Retrodicting the past was impossible in the Copenhagen quantum mechanics of measured subsystems \cite{Har98b}. Retrodiction is possible in a quantum mechanics of closed systems like DH \cite{GH90,Har98b} that provides  probabilities for histories of events  that can extend to the past. In ordinary parlance these are the probabilities that the events {\it happened}. 


Suppose for example that we have data about the universe at the present time $t_p$ represented by a projection $R_{\rm pd}(t_p)$.  We can infer what happened in the past as follows: First specify a set of alternatives histories in the past  $\{C_\alpha\}$  that includes  the present set of alternatives $\{R_{\rm pd}, I- R_{\rm pd}\}$ (cf. \eqref{class-closed}). Assuming this set is recorded  calculate the probabilities of  the alternative past histories  conditioned on thepresent data., viz. 
\be
p(\alpha|{\rm pd})\quad (\text{probabilities for retrodiction}).
\label{pastprobs}
\ee
We can say that given present data  the history $\alpha$ {\it happened} in the past with probability \eqref{pastprobs}. If $R_{\rm pd}$ is one of the records of the set in the sense of \eqref{correlation} then $\alpha$ is certain. But we can not expect to identify such records in cosmology, geology, evolutionary biology and human history. Discussion of the past therefore is probabilistic as a practical matter.

\subsection{Incompatible Pasts} 

There are many different sets of past histories $\{C_\alpha\}$ that could have been chosen to retrodict --- even incompatible ones. 
In quantum theory there is no unique past conditioned on a given present record
\cite{Har98b}.  We cannot say what happens without specifying in which set of past histories it happens. Different past sets of histories can provide
different, even apparently incompatible, stories of what happened consistent with a present record. A striking, if artificial, example of this is
provided by the three-box model introduced by Aharonov and Vaidman for a different purpose \cite{AV91}.



\subsection{The Three Box Model}
\label{threebox}

Consider a particle that can be in one of three boxes, $A$, $B$, $C$ in corresponding
orthogonal states $|A\rangle$, $|B\rangle$, and $|C\rangle$. For simplicity, take the
Hamiltonian to be zero, and suppose the system's initial state to be 
\begin{equation}
|\Psi\rangle\equiv \frac{1}{\sqrt{3}}\ (|A\rangle + |B\rangle + |C\rangle)\, .
\label{threesix}
\end{equation}
Suppose we have a present record represented by 
$R_{\rm pd} = |\Phi\rangle\langle\Phi |$ where 
\begin{equation}
|\Phi\rangle \equiv \frac{1}{\sqrt{3}}\ (|A\rangle + |B\rangle - |C\rangle)\, .
\label{threeseven}
\end{equation}
From this present data we now retrodict different pasts in which different events happen.

Denote the projection operators on $|\Phi\rangle$, $|A\rangle$, $|B\rangle$, $|C\rangle$
by $P_\Phi$, $P_A$, $P_B$, $P_C$ respectively. Denote by $\bar A$ the negation of $A$
(``not in box $A$'') represented by the projection $P_{\bar A}=I-P_A$. The negations
$\bar \Phi$, $\bar B$, $\bar C$ and their projections $P_{\bar\Phi}$, $P_{\bar B}$, and $P_{\bar C}$
are similarly defined.

From the present data $R_{\rm pd} = |\Phi\rangle\langle\Phi |$ and the system's state  $|\Psi\rangle$ let us ask
for the probability that  the particle was in the box $A$ at an earlier time than the present. (The exact values
of the times are unimportant since $H=0$. Only the order matters.) One  past set of histories
is represented by the $\{C_\alpha\}$ 
\begin{equation}
P_\Phi P_A\, ,\ P_\Phi P_{\bar A}\, ,\ P_{\bar\Phi} P_A\, ,\ P_{\bar\Phi} P_{\bar A}\, .
\label{threeeight}
\end{equation}
This set of histories is  easily checked to decohere exactly and therefore is recorded. The conditional probabilities for $A$ and
$\bar A$ given $\Phi$ can be calculated from \eqref{probRH} 
The result is
\begin{equation}
p(A|\Phi)=1\ , \quad p(\bar A|\Phi)=0\, .
\label{threenine}
\end{equation}
Thus, we can say  {\it  in this set of alternative histories of the past}  present data implies that it happened that  the particle was in box $A$. 

An examination of \eqref{threesix} and \eqref{threeseven} shows that both state
and record are symmetric under interchange of $A$ and $B$. Therefore,
using the decoherent set of histories
\begin{equation}
P_\Phi P_B\, ,\ P_\Phi P_{\bar B}\, ,\ P_{\bar \Phi} P_B\, ,\ P_{\bar\Phi} P_{\bar B},
\label{threeten}
\end{equation}
we can compute
\begin{equation}
p(B|\Phi)=1\ , \ p(\bar B|\Phi)=0\, .
\label{threeeleven}
\end{equation}
Thus, we can say {\it in this past set of histories} present data that it happened that the the particle was box $B$. 

There is no contradiction because the sets of histories \eqref{threeeight} and
\eqref{threeten} are incompatible realms. The finer-grained set of histories describing
both $A$ and $B$ is
\begin{equation}
P_\Phi P_A P_B\, , \ P_\Phi P_A P_{\bar B}\, ,\ P_\Phi P_{\bar A} P_B\, , \cdots ,
\ \text{etc.}
\label{threetwelve}
\end{equation}
But this set does not decohere. The inference ``if in $A$ then not in $B$'' cannot be
drawn since there are no consistent probabilities for it.



\subsection{Quantum Physics and Human Language}
\label{language}

There is no conflict with logic or quantum mechanics arising from  the particle in the three box model being in one box in one set of past histories and a different box in another. But there is a conflict with the usual usage of `happen' in human language. 

Human language evolved over tens of thousands of years of human focus one kind of sets of histories --- the quasiclassical realms of everyday experience\footnote{See \cite{Har10} for a review.}. Human languages employ constructions that implicitly assume properties of the limited range of phenomena they evolved to describe. These assumed properties are true features of that limited context, but may not be general 
properties of all the physical situations allowed by fundamental physics as we see with this three box model. 
The surest route to clarity is to express the constructions of human languages in the language of fundamental physical 
theory, not the other way around. In particular, if you say something happened you must also specify in what recorded set of alternative histories it happened 

\section{Comparison of the Formulations}
\label{comparison}

We have exhibited three ways of formulating the decoherent histories quantum mechanics of closed systems like the universe\footnote{ There are more!  Strong decoherence \cite{GH13}, weak decoherence \cite{Gri02, Har04}  linear positivity \cite{GP95}, and quantum logic \cite{Ishsum} for example. However we do not aim at a comprehensive review.}. This section compares them. The next section discusses their possible use. 

\subsection{Three Formulations Summarized}
\label{sumarized}

{\it Original Framework (DH)}:
The original framework assigns probabilities to decoherent sets of alternative histories $\{C_\alpha\}$ for which the quantum interference between the individual members of the set vanishes as quantified by (medium) the decoherence condition 
\be
\label{decoherence}
D(\alpha,\beta) \equiv\langle \Psi_{\alpha}|\Psi_\beta\rangle = \langle\Psi |C_\alpha^\dagger C_\beta |\Psi\rangle \approx 0, \quad  \alpha\ne\beta,  \quad \quad \text{(decoherence)}.
\ee
Sets of  histories satisfying this condition are said to {\it decohere.} The decoherence condition \eqref{decoherence} captures in a general and abstract way realistic physical mechanisms leading to the decoherence of histories\footnote{For example see \cite{GH93,BH99} for some of the author's calculations. }.
The probabilities for the individual histories in a decoherent set are 
\be
\label{prob1}
p(\alpha) = || |\Psi_\alpha \rangle||^2 \approx ||C_\alpha|\Psi||^2.
\ee
These are consistent with the rules of probability as a consequence of the decoherence condition \eqref{decoherence} and \eqref{exhaustC}

We can reexpress the probabilities \eqref{prob1} using the following relation which is a consequence of the decoherence condition  \eqref{decoherence} and \eqref{exhaustC}
\be
\label{summingout}
\langle\Psi|C_\alpha|\Psi\rangle = \sum_\beta  \langle \Psi|C_\beta^\dagger C_\alpha |\Psi\rangle \approx 
\langle\Psi|C_\alpha^\dagger C_\alpha |\Psi\rangle .
\ee
Then, using \eqref{prob1}, 
\be
\label{probDH}
p(\alpha) \approx \Re \langle\Psi|C_\alpha|\Psi\rangle. 
\ee

{\it Recorded Histories (RH):} This is the formulation of DH given in this paper so it hardly needs a summary. RH assigns  probabilities to recorded histories 

{\it Extended Probabilities (EP):}  
In quantum theory there are alternatives which  can be described but that are not the basis for settleable bets. The two-slit experiment in Figure 1 provides an example in the context of the approximate (Copenhagen) quantum mechanics of measured subsystems.  An electron starts at a source, passes through a screen with two slits, and is detected at a point $y$ on a further screen. Consider the two alternative histories distinguished by whether the electron went through the upper slit or the lower slit to arrive at a given point $y$. 

If a measurement determines which slit the electron passed through, quantum mechanics provides probabilities for a bet  on which slit the electron went through. A record of the measurement outcome can be used to settle the bet. If no measurement is carried out, the alternative histories going through the upper and lower slit can still  be described.  But, because of quantum interference, there can't be a record of which slit the electron passed through because then there would be no interference pattern\footnote{For a beautiful example of the transition between not recorded and recorded in an interference experiment see \cite{hotbucky}.}.    
A bet on these alternative histories is not settleable. 
\begin{figure}
\begin{center}
\includegraphics[width=3in]{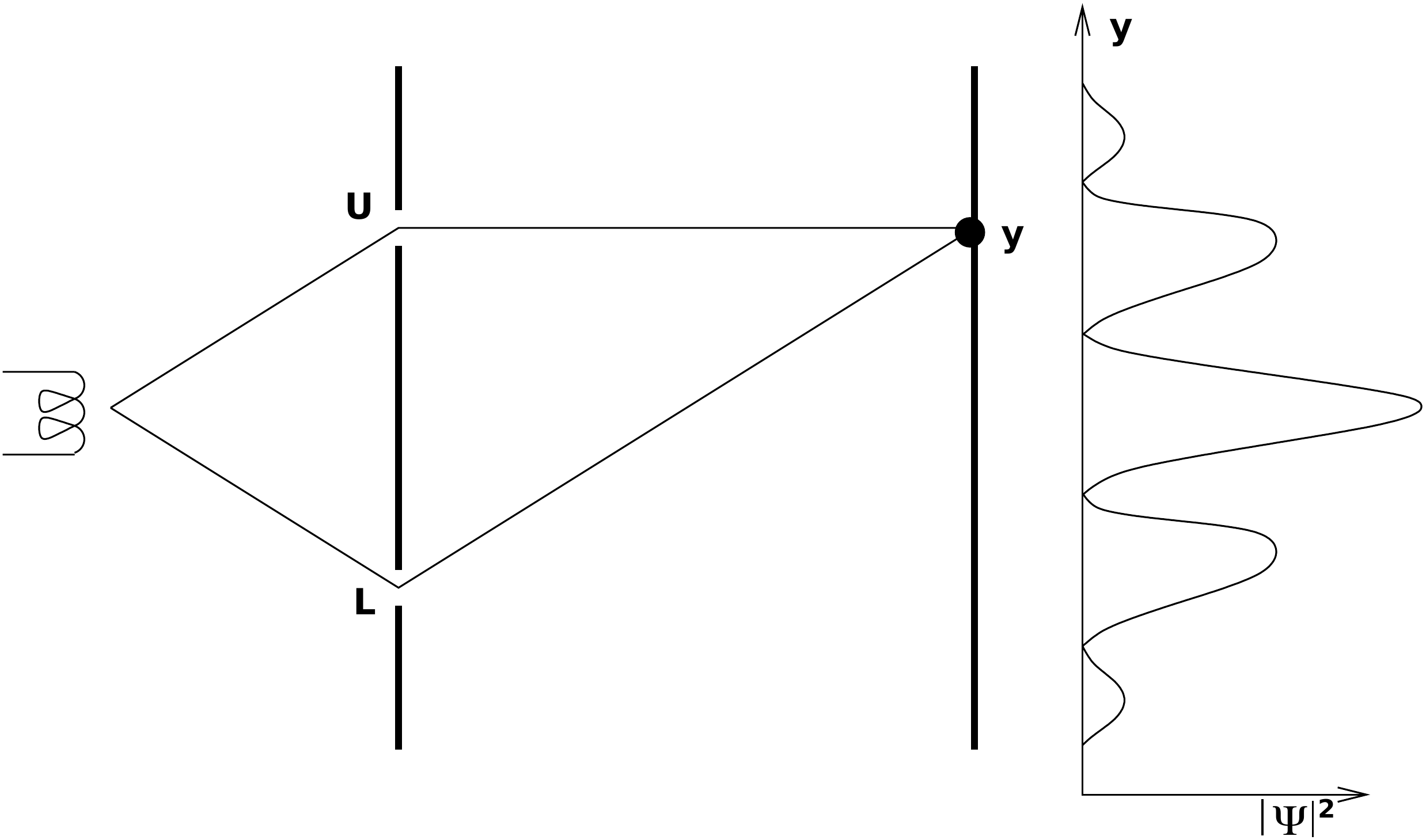}
\caption{The two-slit experiment. An electron gun at left emits an electron
traveling towards a screen with two slits, its progress in space recapitulating
its evolution in time. The electron is detected at a further screen at a position $y$ with a probability density that exhibits an interference pattern. A coarse grained set of histories for the electron is defined by specifying the slit ($U$ or $L$) through which the electron passes through and ranges $\Delta$ of the position $y$ where it is detected. In the absence of the record of  a measurement it is not possible to settle a bet on the which of these histories occurred.}   
\label{2slit}
\end{center}
\end{figure}

When there are alternatives that can be described but do not correspond to settleable bets there are two approaches to probability.   (1) Assign probabilities only to alternatives that  correspond to settleable bets in which case the usual rules of probability theory follow. (2) Assign probabilities to all alternatives, settleable or not, but allow for extensions of the usual probability theory rules for non-settleable bets. The first approach is the one taken in DH.  EP explores the second option.

 EP  \cite{GH13} starts from the formula
\be
\label{extended}
\wp(\alpha) \approx \Re \langle\Psi|C_\alpha|\Psi\rangle. 
\ee
This equation does not define a set of probabilities because the right hand side does not always lie between zero and one. 
Rather \eqref{extended} defines {\it extended } probabilities that satisfy all the rules of probability theory except lying between zero and one.  A set with extended probabilities outside the zero-one range can never be recorded in the sense of  \eqref{correlation} because the relation \eqref{probRH} could not be satisfied. A set with out of range zero to one probabilities can be understood as an instruction not to bet on it because there will be no records to settle the bet\footnote{For more on the consequences of extended probabilities see \cite{HY12}.}.

When all the $\wp(\alpha)$ happen to be in range than the $\wp(\alpha)$ define probabilities $p(\alpha) =\wp(\alpha)$ which from \eqref{extended} are given by
\be
\label{probEP}
p(\alpha) \approx \Re \langle\Psi|C_\alpha|\Psi\rangle. 
\ee

\subsection{Three Formulations of Quantum Mechanics Compared}

In all three formulations the probabilities for the histories $\{C_\alpha\}$ are given by the same formula viz. $p(\alpha) \approx \Re \langle\Psi|C_\alpha|\Psi\rangle$ (cf.\eqref{probRH},\eqref{probDH},\eqref{probEP}).
That is why we can call the three cases different formulations of the same theory---DH. The three formulations differ in the conditions under which the formula holds. 

All three formulations are equivalent under the assumption of strong records \eqref{strong}. Strong records imply decoherence, imply that histories are recorded (obviously), and imply that the extended probabilities are in fact probabilities obeying the usual rules of probability theory. 

If strong records are not assumed then the sequence EP, RH, DH is a progression from formulations with fewer assumptions to ones with more --- from weaker to stronger. One can see that explicitly from the status of records. In  EP records need not exist to have probabilities. In RH they are required as an assumption, and in DH they are automatic and a consequence of  the assumption of  decoherence.  

There are recorded histories that are not decoherent. Histories are recorded if the branch state vectors $|\Psi_\alpha\rangle \equiv C_\alpha|\Psi\rangle$ lie in orthogonal subspaces, but decoherence requires the branch state vectors to be orthogonal. 

Of what use are these weaker formulations? We describe a possible answer in the next section.

\section{Extensions and Generalizations of DH}
\label{generalizations}
As already mentioned, decoherent histories quantum mechanics (DH) abstracts and incorporates many realistic features of the world of laboratory science where quantum mechanics has been tested and many beautiful quantum phenomena realized. On `macroscopic scales' the intuition of some prominent scientists is that the principle of superposition will have to be modified \cite{qmmod} and with it the rest of the laws of quantum mechanics. On much larger scales  we hope to extend quantum mechancis  to the very early universe where spacetime geometry is not fixed as is assumed in DH but fluctuating quantum mechanically and without definite value.
It therefore  seems likely that DH will have to be generalized or modified further. Different formulations of DH provide different starting points for making these generalizations.  That is why they are of interest. 

In the following we describe two of these situations in a bit more detail:

\subsection{Quantum Spacetime}

\label{quantspacetime}

DH as presented here assumes a fixed background spacetime geometry.  A fixed background spacetime geometry is approximately true in the late universe where we live ---approximately 14Gyr from the big bang. But in the very early universe, near the big bang, we expect spacetime geometry to be fluctuating quantum mechanically.  The evidence for such fluctuations is all about us in the CMB temperature fluctuations, the large scale structure  in the today's distribution of the galaxies, and the gravitational wave signals that we may see in CMB polarization.  We need a generalization of DH that does not require a fixed notion of time but can deal with spacetime as a quantum variable.

Generalized quantum mechanics \cite{Har91a,Ishsum} is a framework abstracting the ideas of DH  in which its possible to look for generalizations adequate for quantum spacetime.  An example is the sum-over-histories generalized quantum mechanics of semiclassical quantum spacetime (e.g. \cite{Har95c,Har07a}).  But alternative generalizations of quantum mechanics may be produced starting from the other formulations of DH that have been discussed in this paper. 

\subsection{Testing the Theory}
\label{testing}
To aid the experimental search for deviations from DH  it would be very useful to have theories that are close to DH but not DH itself (to paraphrase Weinberg \cite{Wei92}). Alternative formulations of DH could be starting points for this goal. 

\subsection{Emergent Quantum Mechanics?} 
\label{emergent}
In the history of physics ideas that were once accepted as fundamental, general, and inescapable were later seen to be consequent, special, and dispensable. These ideas were not truly a general feature of the world, but only seemed to be general because of  our special situation in the universe and the limited range of our experience. They were excess baggage that needed to be discarded to reach a more general perspective \cite{Har90b}. Could some features of DH be excess baggage?  In  applying quantum mechanics to the whole universe for instance why do we need the principle of superposition when the universe has a single quantum state?  We expect classical spacetime to be an emergent feature of the world in a quantum theory of gravity. But could DH, with its assumption of classical spacetime, also emerge along with classical spacetime from something deeper \cite{Har07a}.  Various ways of formulating quantum DH will help us find out. 

\section{Conclusion}
\label{conclusion}

Alternative formulations of DH such as those discussed in this paper  can be useful both for understanding the theory and for extending it to new realms of application and experimental test. 

\acknowledgments  

The author thanks Murray Gell-Mann, Thomas Hertog, and Mark Srednicki  for discussions on the quantum mechanics of the universe over long periods of time. He thanks the Santa Fe Institute for supporting many productive visits there. The this work was supported in part by the National Science Foundation under grant PHY15-04541.


\end{document}